\documentclass[a4paper,twocolumn,11pt,accepted=2021-08-05]{quantumarticle}
\pdfoutput=1

\usepackage{amsfonts, amsmath, amssymb}
\usepackage[utf8]{inputenc}
\usepackage[english]{babel}
\usepackage[T1]{fontenc}
\usepackage{braket}
\usepackage{hyperref}
\usepackage{bm}
\usepackage{color}
\usepackage{soul}
\usepackage{mathtools}
\usepackage{float}
\usepackage{ulem}

\usepackage[backend=bibtex,sorting=none]{biblatex}
\newcommand{\be}{\begin{equation}}
\newcommand{\ee}{\end{equation}}
\newcommand{\ba}{\begin{eqnarray}}
\newcommand{\ea}{\end{eqnarray}}
\newcommand{\h}{\hat{\mathcal{H}}}

\newcommand{\ssp}{\hat{\sigma}^+}
\newcommand{\ssm}{\hat{\sigma}^-}
\newcommand{\ssx}{\hat{\sigma}^x}

\newcommand{\ssz}{\hat{\sigma}^z}
\newcommand{\ssa}{\hat{\sigma}^\alpha}

\newcommand{\pmeas}{p_{\rm meas}}
\newcommand{\psite}{p_{\rm site}}
\newcommand{\mie}{\Theta_E}

  {\left\lbrace\begin{array}{@{}l@{}}}%
  {\end{array}\right.}

\newcommand{\blank}{\vspace{3mm}\noindent}

\begin{document}

\title{Many-body quantum Zeno effect and measurement-induced subradiance transition}

\author{Alberto Biella}
\email{alberto.biella@universite-paris-saclay.fr}
\affiliation{Universit\'e Paris-Saclay, CNRS, LPTMS, 91405 Orsay, France}
\affiliation{INO-CNR BEC Center and Dipartimento di Fisica, Universit\`a di Trento, 38123 Povo, Italy}
\affiliation{JEIP, USR 3573 CNRS, Coll\`{e}ge de France, PSL Research University, 11 Place Marcelin Berthelot, 75321 Paris Cedex 05, France}

\author{Marco Schir\'o}
\email{marco.schiro@college-de-france.fr}\thanks{on leave from: Institut de Physique Th\'{e}orique, Universit\'{e} Paris Saclay, CNRS, CEA, F-91191 Gif-sur-Yvette, France}
\affiliation{JEIP, USR 3573 CNRS, Coll\`{e}ge de France, PSL Research University, 11 Place Marcelin Berthelot, 75321 Paris Cedex 05, France}

\begin{abstract}
It is well known that by repeatedly measuring a quantum system it is possible to completely freeze its dynamics into a well defined state, a signature of the quantum Zeno effect. Here we show that for a many-body system evolving under competing unitary evolution and variable-strength measurements the onset of the Zeno effect takes the form of a sharp phase transition. Using the Quantum Ising chain with continuous monitoring of the transverse magnetization as paradigmatic example we show that for weak measurements the entanglement produced by the unitary dynamics remains protected, and actually enhanced by the monitoring, while only above a certain threshold the system is sharply brought into an uncorrelated Zeno state. We show that this transition is invisible to the average dynamics, but encoded in the rare fluctuations of the stochastic measurement process, which we show to be perfectly captured by a non-Hermitian Hamiltonian which takes the form of a Quantum Ising model in an imaginary valued transverse field.  We provide analytical results based on the fermionization of the non-Hermitian Hamiltonian in supports of our exact numerical calculations. 
\end{abstract}
\maketitle

\begin{figure}[b!]
\centering
\includegraphics[width=.9\columnwidth]{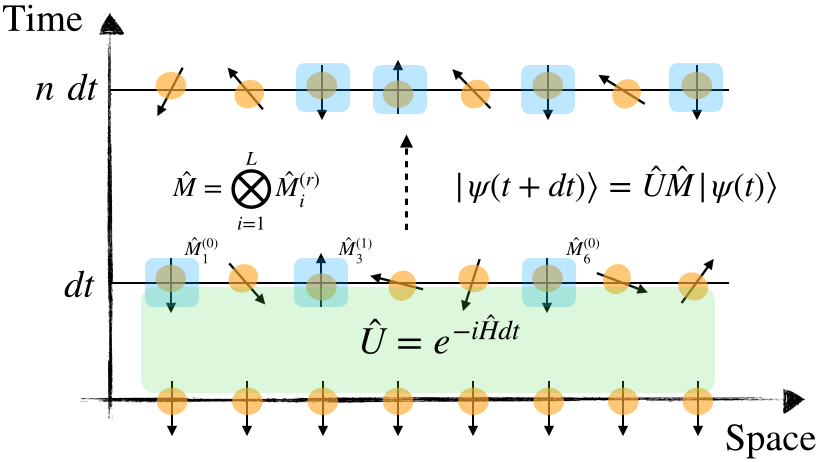}
\caption{Cartoon of the hybrid quantum dynamics considered in this work [see Eq.~(\ref{Measurements})] where in addition to the unitary evolution $\hat{U}$ (in our example generated by the Quantum Ising Model) there is a stochastic measurement process on each site $\hat{M}=\bigotimes_{i=1}^L \hat{M}^{(r)}_i$.}
  \label{fig:sketch}
\end{figure}

\section{Introduction}
Measurements lie at the heart of quantum mechanics and of its intrinsic statistical interpretation~\cite{wiseman_milburn_2009}. They are known to fundamentally perturb the state of a quantum system, leading to the famous wave-function collapse, or to hamper its evolution in the Hilbert space, as in the celebrated Quantum Zeno Effect (QZE), where frequent projective measurements on a system can lead to its full localization~\cite{misra1977thezeno,peres80zeno,itano1999quantum,facchi2001from,facchi2002quantum,signoles2014confined,
snizhko2020quantumzeno}. In the field of quantum information the role of measurements as a resource has been extensively discussed and also experimentally demonstrated~\cite{duan2001longdistance,sorensen2003measurement,chou2005measurement,roch2014observation,
kong2020measurement}.
In the context of quantum many-body systems the interest is much more recent. Manifestations of many-body QZE have recently attracted interest~\cite{syassen2008strong,patil2015measurement,froml2019fluctuation,froml2020ultracold,krapivsky2019free,krapivsky2020free}. A series of theoretical works have explored the interplay between entangling quantum many-body dynamics and frequent projective measurements, giving rise to novel exotic entanglement phase transitions~\cite{Li2018quantum,skinner2019measurement,szyniszewski2019entanglement,choi2020quantumerror,jian2020measurement,
Turkeshi2020measurement}. A different measurement protocol is provided by a continuous monitoring, or variable strength weak measurements, which is defined in terms of positive operator-valued measurements (POVM) and provides a stochastic backaction onto the system~\cite{bernard2018transport,cao2019entanglement,ivanon2020feedback,yang2020quantum}. Also in this case entanglement phase transitions have been identified~\cite{szyniszewski2019entanglement,fuji2020measurement}. A crucial feature of these transitions is that they are invisible to the average dynamics but only appears at the level of single quantum many-body states.

In this work we address the competition between unitary many-body dynamics and stochastic quantum measurements in a simple and paradigmatic setting, namely when the two processes involve degrees of freedom which do not mutually commute, such as different components of a quantum spin. Surprisingly we find, in the context of a one-dimensional quantum Ising model with continuous monitoring of the local transverse magnetization, that the onset of the many-body QZE, in which the system is freezed by measurements into a well defined many-body state, appears sharply above a critical measurement strength, a behavior suggestive of a novel form of quantum criticality. As for the above mentioned entanglement phase transitions, we show here that this sharp transition is encoded in the rare realizations of the stochastic process and cannot be observed by only looking at the average behavior of the system. We highlight this point by considering  the no-click evolution,  corresponding to a post-selection of the rare {\it no-click} events, which can be described by a non-Hermitian Hamiltonian~\cite{ashida2020nonhermitian}, which in our case takes the form of a non-Hermitian Transverse Field Ising Model (TFIM)~\cite{lee2014heralded}. Remarkably we show that the sharp onset of the many body QZE can be understood in terms of a subradiance transition~\cite{dicke1954coherence,gross1982superradiance,celardo2009superradiance,auerbach2011superradiant,biella2013subradiant,
guerin2016subradiance,rotter2015areview} in the many-body spectrum of the non-Hermitian TFIM, i.e. a non-analytic change in the eigenstate with the smallest imaginary part. This result suggests an intriguing analogy with the concept of a quantum phase transition, where the system ground-state properties change abruptly as a result of the competition between non-commuting operators.  We provide analytical arguments based on the fermionization of the non-Hermitian TFIM in supports of our exact numerical calculations. 
We stress that the many-body nature of the problem is encoded in the competition between unitary entangling dynamics and local measurements, rather than in the unmeasured system itself (the quantum Ising chain) which is integrable through free fermions techniques.
We also note that to the best of our knowledge this mapping cannot be used to exactly solve the problem for the measurement protocol considered in this work, corresponding to a Born rule~\cite{nahum2021measurement}, as opposed to other stochastic settings~\cite{xhek2021,Alberton2021} which also show genuine many-body signatures such as entanglement phase transitions.

\section{The model and the measurement protocol }
In this work we consider a stochastic quantum many-body dynamics governed by the following protocol 
\be
\label{protocol}
\ket{\psi(t+dt)} = \hat{M} \hat{U}\ket{\psi(t)},
\ee
where the unitary dynamics $\hat{U}=\exp(-i \h dt)$ is generated by a Quantum Ising model, 
\be\label{HIsing}
\h=J^x\sum_{i=1}^{L-1}\ssx_i\ssx_{i+1}
\ee
 ($\ssa_i, \alpha=x,y,z$ being the Pauli matrices acting on the $i$-th site) and $\hat{M}$ is a measurement carried out simultaneously on each site with probability $\psite$. The operator $\hat{M}$ is a (normalized) Kraus operator associated with a POVM. Specifically we have
$
\hat{M}=\bigotimes_{i=1}^L \hat{M}_	i^{(r)}, 
$
with $r=0,1$ defined by
\ba\label{Measurements}
\hat{M}_i^{(0)} &=& \hat{\Pi}^z_{i-}+\sqrt{1-\pmeas \ p_{\rm site}}\,\hat{\Pi}^z_{i+},\\
&&\cr
\hat{M}_i^{(1)} &=& \sqrt{\pmeas \  p_{\rm site}}\,\hat{\Pi}^z_{i+},
\ea
where $\hat{\Pi}^z_{i\pm}=\left(\mathbb{I}\pm\ssz_i\right)/2$ are local projectors onto eigenstates of the spin $z$-component $\ssz_i\ket{\pm}_i=\pm \ket{\pm}_i$.  Locally, the probabilities of the two possible readouts are given by
$p^{(1)}_i = \pmeas\, p_{\rm site}  \langle\Psi(t)\vert\hat{\Pi}^z_{i+}\vert\Psi(t)\rangle, \quad p^{(0)}_i=1-p^{(1)}_i$.

Equations (\ref{protocol}-\ref{Measurements}), together with the initial condition that we fix to be an uncorrelated state with all the spin aligned along the $z-$axis i.e. $\vert\Psi(0)\rangle=\bigotimes_{i=1}^L \ket{-}_i$, fully specifies the stochastic protocol we will focus on in this work (See Figure~\ref{fig:sketch} for a sketch of the evolution). 
The choice  of the initial state is rather {\it natural} in this context. 
Starting from an uncorrelated product state we want to study how the non local unitary evolution entangles different sites and local measurements kill correlations.
This is alternative to a scenario involving non-local measurement processes which is expected to stabilize correlated entangled states.
While this problem is is in principle treatable theoretically it would be very demanding to implement experimentally.

Each random realization provides a quantum many-body state encoded in the wave function $\vert\Psi(t)\rangle$. We stress that Eq.~\eqref{protocol} leads to an unnormalized wavefunction since $\hat{M}$ is not a unitary operator. We thus normalize the quantum state after each step of the protocol.

\section{Results}
In the following we are going to discuss the stochastic dynamics of the system as a function of the ratio between the frequency of the measurement and the unitary dynamics. It is natural therefore to introduce the adimensional control parameter
\be
\label{cpar}
g=\frac{\pmeas}{J^x \ dt},
\ee 
which will play a key role in the following. Here the parameter $dt$ is finite and defined through the unitary evolution in Eq.~\eqref{protocol}. The physics in two limiting cases is easy to grasp. For $g\ll 1 $, quantum jumps (the occurrence of $\hat{M}_i^{(1)}$) take place with a very low probability ($p_{r=1}\ll1$) regardless of the value of $\psite$ and the projective measurement  $\hat{M}_i^{(0)}$ is close to the identity. The system therefore evolves under the unitary dynamics with rare measurements. In the opposite limit $g\gg1$ the measurements occur often with respect to the typical timescale settled by the unitary dynamics. In this regime the stochastic evolution starting from our initial condition stabilizes the uncorrelated state $\ket{T}=\bigotimes_{i=1}^L \ket{-}_i$ which is also the exact fixed point of the measurement protocol in the limiting case $\pmeas=1$. As we are going to show below those two regimes are not smoothly connected, but a sharp phase transition between two well defined phases occur upon increasing $g$. Importantly, this will appear in the rare fluctuations but not in the average dynamics that instead display a rather featureless and smooth crossover.

\subsection{Measurement-induced entanglement}
We start by studying the entanglement properties encoded in 
the stochastic evolution of a single quantum state, as a function of $g$.  Specifically, we compute the entanglement entropy of a symmetric bipartition of the system
\begin{figure}[b!]
\centering
\includegraphics[width=0.9\columnwidth]{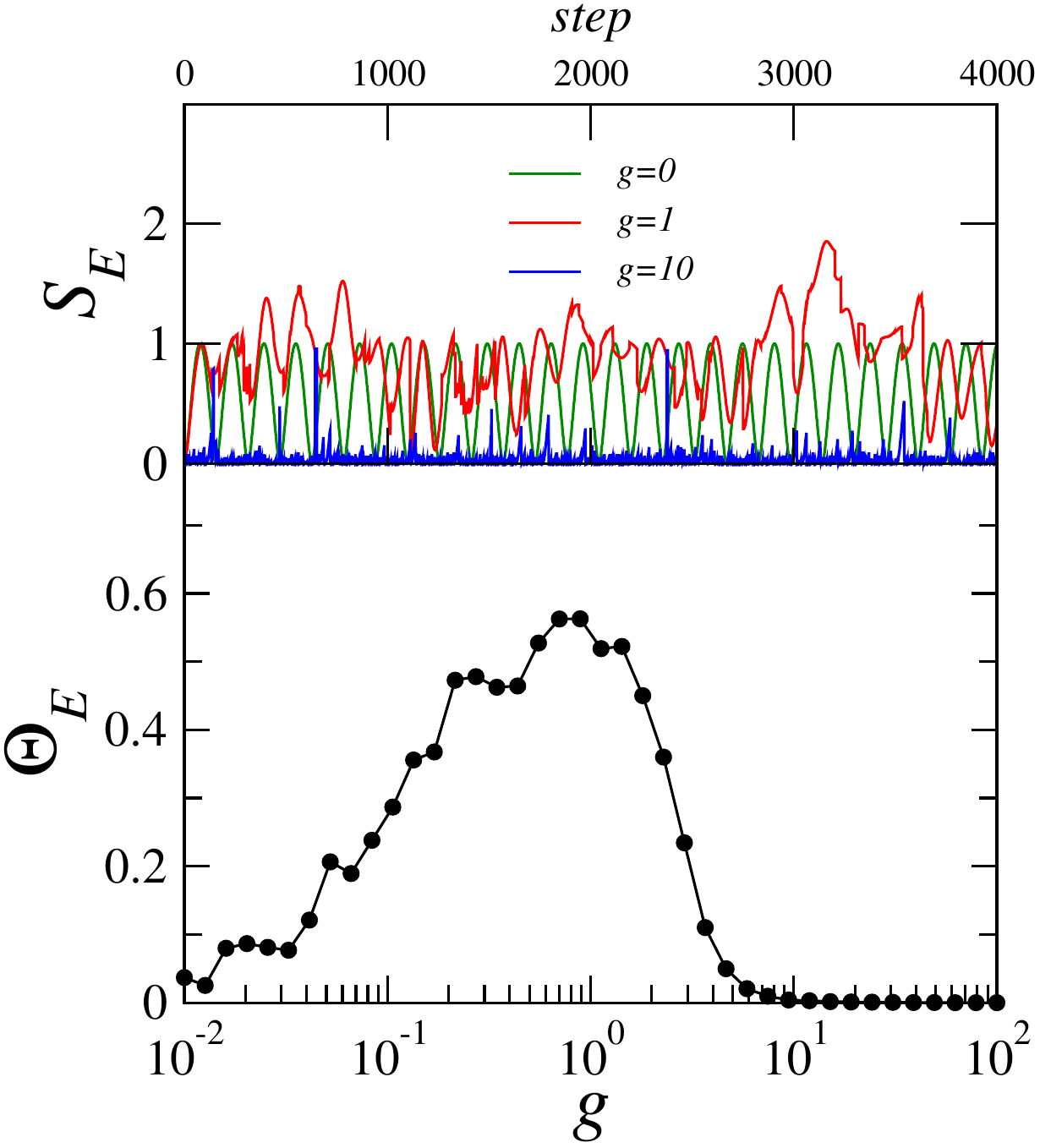}
\caption{Top panel: Evolution of the entanglement entropy $S_E(t)$ of a single quantum state for different values of $g$.  
Bottom panel: measurement-induced entanglement $\mie$ as a function of $g$.
Here we set $N_{\rm real}=100, dt=10^{-2},L=8,J^x=1$.}
  \label{fig:MIEE}
\end{figure}
\be
S_E(t) = -{\rm Tr}\left[\hat{\rho}_A(t) \log_{2}\hat{\rho}_A(t) \right], 
\ee
where $\hat{\rho}_A(t)={\rm Tr}_B\left[\ket{\psi(t)}\bra{\psi(t)}\right]$ is the reduced density matrix of the left halve (labelled as $A$) of the system. In the purely unitary (and deterministic) case, $g=0$, this quantity displays coherent Rabi oscillations in the range $0\le S_E\le1$ (see Fig.~\ref{fig:MIEE}, top panel). Remarkably, we find that for small finite $g$ measurements can produce {\itshape more entanglement} between the two halves of the chain with respect to the unitary case, as shown by the  entanglement entropy of a single stochastic realization, going above $S_E=1$. For large $g$ instead, as expected, the entanglement production is strongly suppressed.
In order to quantify this {\it entanglement generation} with respect to the unitary case ($g=0$)  we consider the probability $P(S_E)$ that a certain value of $S_E$ occurs during the stochastic evolution and then we evaluate the measurement-induced entanglement $\mie$
\be
\label{mie}
\mie = \int_{S_E>1} S_E P(S_E) \ dS_E,
\ee
which is a measure of the entanglement generated on top of the unitary dynamics averaged over $N_{\rm real}$ stochastic realizations.
In Fig.\ref{fig:MIEE} (bottom panel) we show the behavior of $\mie$ as a function of $g$ and for $p_{\rm site}=1$. We find a surprising non-monotonic behavior of the excess entanglement as the strength of the measurements is increased, with a maximum around $g\sim 1$ and a rapid decrease for larger value of $g$, ultimately expected as the QZE kicks in and the system is more and more projected towards the product state $\vert T\rangle$. A similar behavior is found for $p_{\rm site}\neq 1$. In fact we find evidence for a data collapse suggesting the excess entanglement $\mie$ only depends on the product $g \ p_{\rm site}$ (see App.~\ref{app:miee}) This result shows that before the onset of the many body QZE the system at the single trajectory level is highly entangled, due to the interplay between unitary dynamics and measurements. As we are going to discuss below this result is the consequence of a transition between two phases, very different in nature, controlled by $g$.

\begin{figure}[t!]
\centering
\includegraphics[width=0.9\columnwidth]{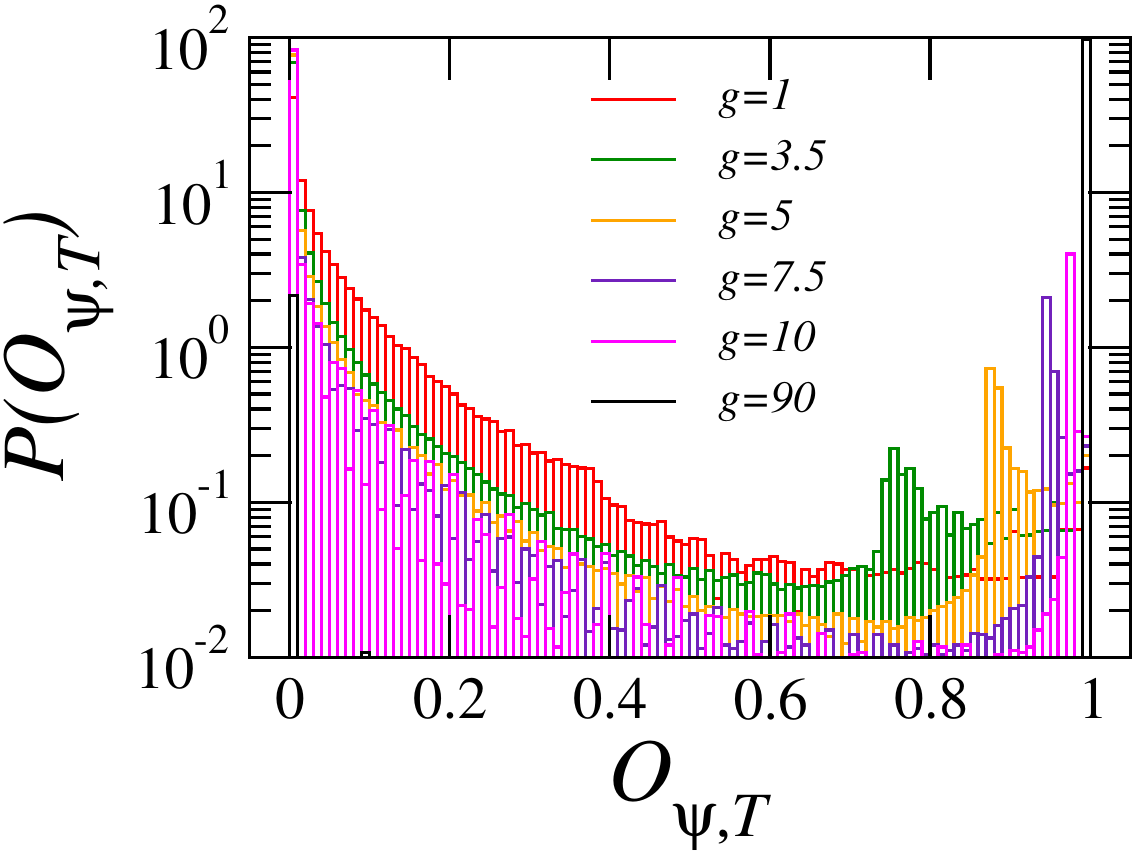}
\includegraphics[width=0.9\columnwidth]{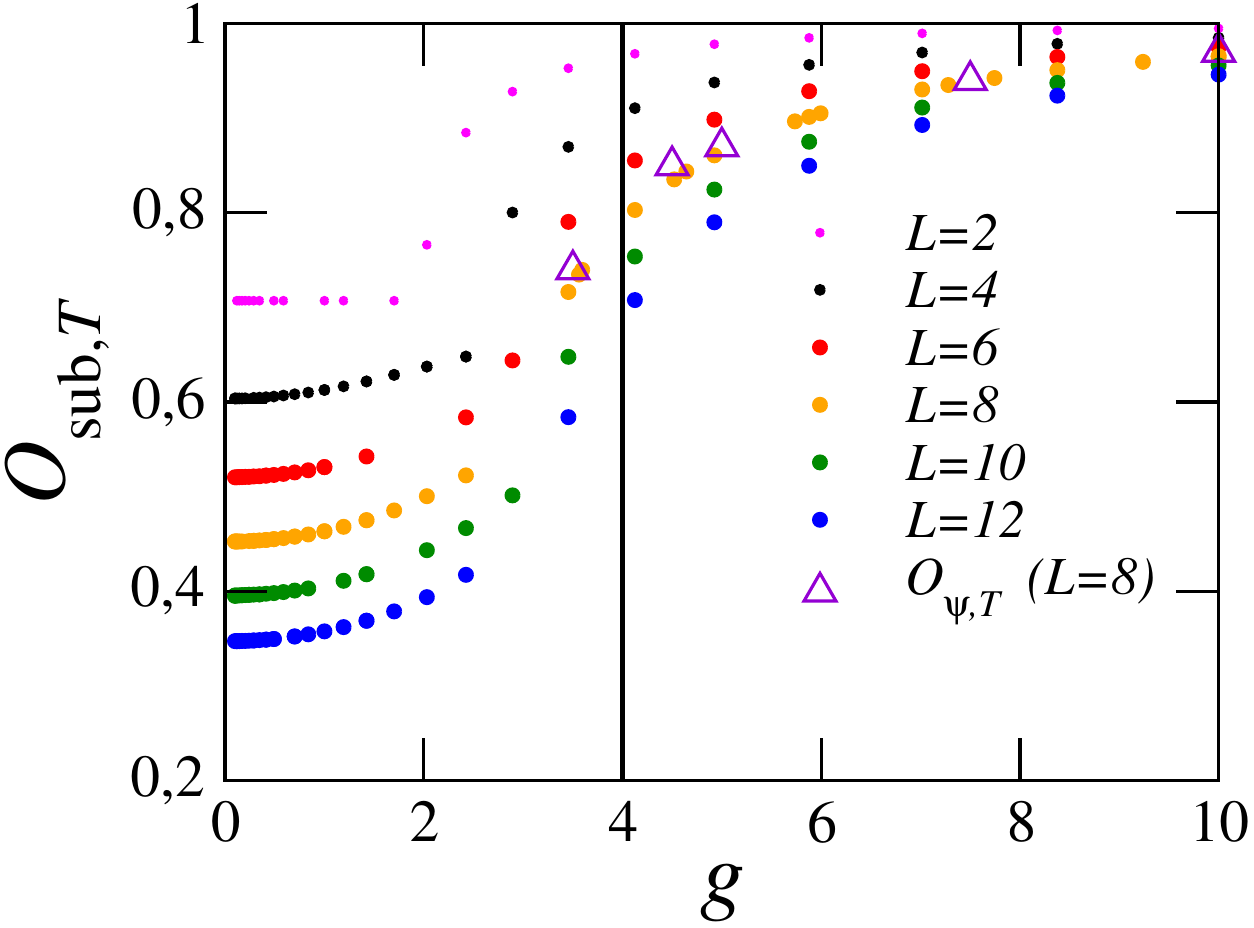}
\centering
\caption{Top Panel: Probability distribution of the overlap $O_{\psi, T}=\left|\braket{\psi | T}\right|$ between the long-time state of the stochastic dynamics in Eq.~(\ref{protocol}) and the Zeno state $\vert T\rangle$. Increasing the strength of the measurements $g$ the distribution develops a secondary peak, which becomes dominant for large $g$.  Here we set $N_{\rm real}=100, dt=10^{-2},L=8,J^x=1$.
Bottom Panel: Overlap between the long-time state of the non-Hermitian evolution and the Zeno state $\vert T\rangle$ as a function of $g$ and for different sizes $L$. For $L=8$ we compare this overlap with the result obtained in the full dynamics by tracking the {\itshape secondary} peak (triangles) and find excellent agreement for $g>g_c$.
} \label{fig:Poverlap}
\end{figure}


\subsection{Onset of many-body quantum Zeno effect}
As already discussed, for $g\gg 1$ the measurements drive the system towards the uncorrelated product state $\vert T\rangle$. This is a signature of the QZE which appears in the stochastic dynamics of local observables, such as the transverse magnetization $m_z$ whose probability distribution features a rich evolution (see App.~\ref{app:zeno}).

Here we discuss the onset of QZE in the probability distribution of the overlap $O_{\psi, T}=\left|\braket{\psi | T}\right|$ at long times, that we plot in Fig.~\ref{fig:Poverlap} (top panel) for different values of $g$.  For $g\lesssim 1$ the function $P(O_{\psi, T})$ is peaked at  $O_{\psi, T}=0$, signaling the correlated nature of $\ket{\psi(t)}$ in the regime where the unitary quantum dynamics dominates, but display a skewed distribution with fat tails slowly decaying for larger value of $O_{\psi, T}$. The broad statistics of $O_{\psi, T}$ suggests that taking the average overlap as measure of the onset of the QZE misses important aspects of the physics, and that a-typical values of the overlap might play an important role in the stochastic dynamics.  Interestingly, for $g\gtrsim 1$ a second peak emerges at  $O_{\psi, T}>0$ and the distribution acquires a strong bi-modal character. 
As $g$ is increased further into the deep quantum Zeno regime this secondary peak (i) shifts towards values of order one, $O_{\psi, T}\sim 1$, indicating that the long-time limit state resembles more and more the uncorrelated product state $\vert T\rangle$, and (ii) acquires larger weight eventually becoming the dominant peak of the distribution. 
Next, we show that the sharp onset of this secondary peak in the overlap distribution can be understood by considering 
the no-click evolution, corresponding to a post-selection of the rare dynamics with outcome $r=0$ for all the lattice sites.

\begin{figure}[h!]
\centering
\includegraphics[width=0.9\columnwidth]{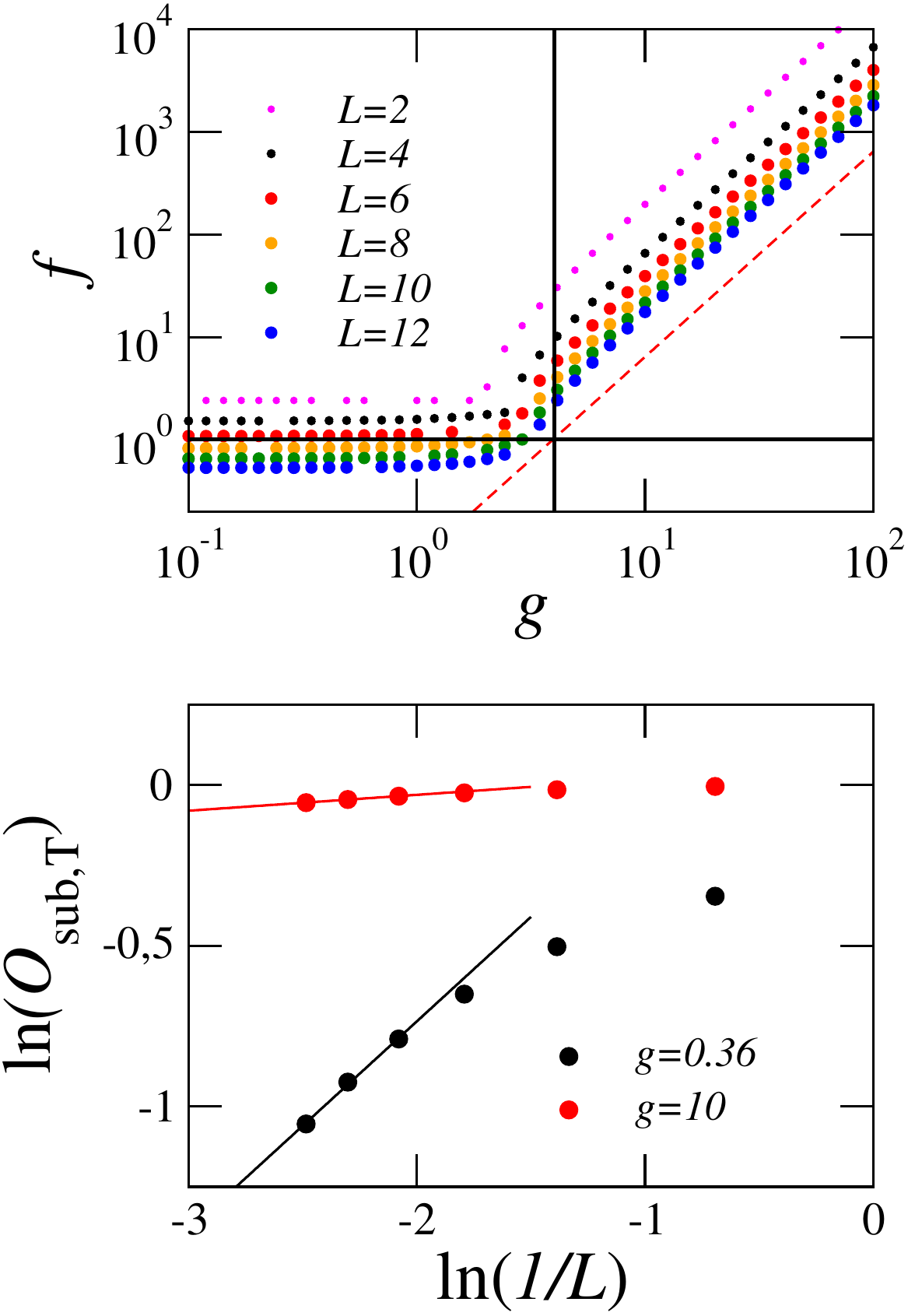}
\caption{
Top Panel: Ratio between the probability that the subradiant state is inside the Zeno subspace and its complement [$f=O_{{\rm sub}, T}/\left(1-O_{{\rm sub}, T}\right)$] as a function of $g$, showing two clearly distinct behaviors. At small $g$ this ratio is constant and small, $f<1$ decreasing with $L$, while for large $g$ we find $f\sim g^2$. 
Bottom panel: Scaling with system size $L$ of the overlap $O_{{\rm sub},T}$ between the subradiant state and the Zeno state $\vert T\rangle$, for two values of $g$. While at large $g$ the overlap depends weakly on $L$, we find indication for a power-law scaling at small $g$. Here we set $dt=10^{-2},J^x=1$.
}
  \label{fig:Sub_size2}
\end{figure}

\subsection{No-click dynamics and subradiance transition}
In absence of a click the POVM reduces to $\hat{M}=\bigotimes_{i=1}^L \hat{M}_i^{(0)}$ regardless of the value of the system parameters.  
Fixing $\psite=1$ and parametrizing the measurement rate as $\pmeas=\gamma dt\rightarrow 0$, we obtain that the dynamics can be described by an effective non-Hermitian evolution (see App.~\ref{app:heff_validity} for a discussion of its validity). 
We thus get $\ket{\psi(t+dt)} \simeq e^{-i \h_{\rm eff} dt}\ket{\psi(t)}$, 
where we have introduced the non-hermitian Hamiltonian
\be
\label{hdissdef}
\h_{\rm eff} = \h -i \hat{\Gamma}= J^x\sum_{i=1}^{L-1}\ssx_i\ssx_{i+1}-i\frac\gamma2\sum_{i=1}^L\frac{\ssz_i+\mathbb{I}}{2}.
\ee
The continuous monitoring generates a purely imaginary transverse field, proportional to the strength of the measurement $\gamma$, which does not commute with the Hamiltonian $\h $. The no-click dynamics is therefore described by a TFIM with a purely imaginary transverse field.

\begin{figure*}[t!]
\centering
\includegraphics[width=1.75\columnwidth]{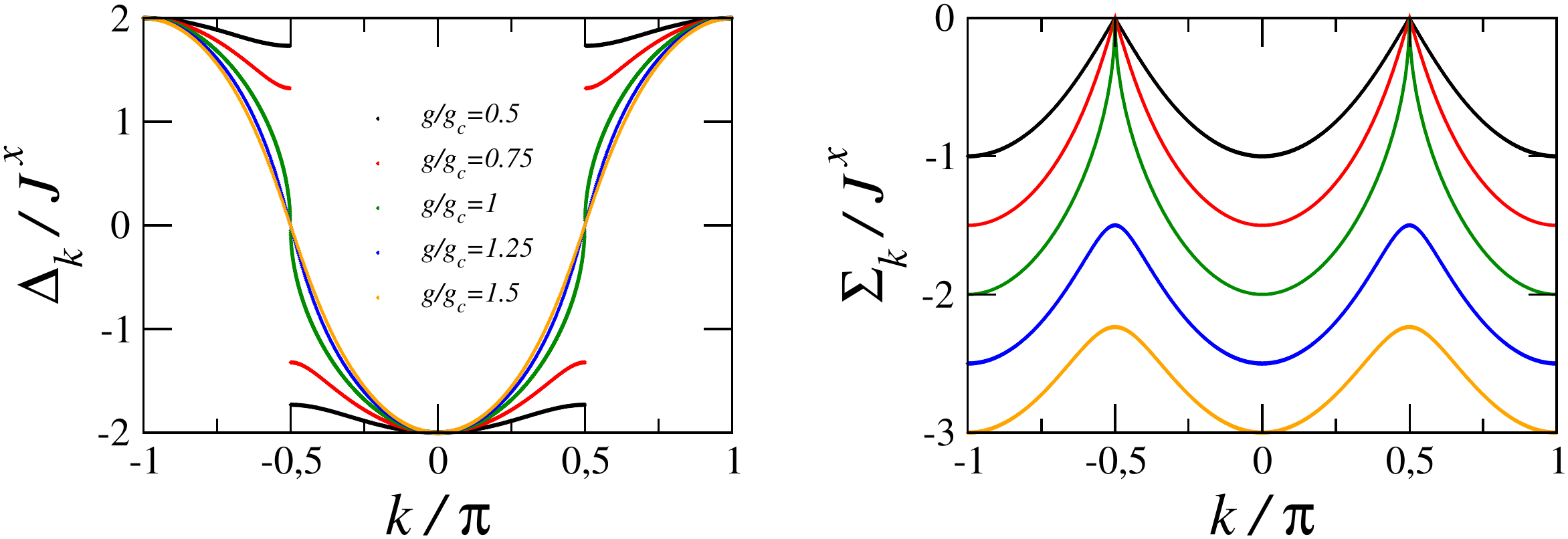}
\caption{Quasiparticle spectrum $\Lambda_k$ [given in Eq.~(\ref{eq:lambdak})] of the non-Hermitian TFIM for different values of $g=\gamma/J^x$. 
For convenience, we choose the sign
convention for each $k$ such that the imaginary part of $\Lambda_k$ is
always negative~\cite{lee2014heralded}.
The real part (left panel), ${\rm Re}\left[\Lambda_k\right]=\Delta_k$, is gapped at $k=\pm\pi/2$ for $g<g_c$ while becomes gapless for $g>g_c$.
The imaginary part (right panel), ${\rm Im}\left[\Lambda_k\right]=\Sigma_k$, always displays two maxima located at $k=\pm \pi/2$. For $g<g_c$ we have $\Sigma_{k=\pm \pi/2}=0$ while for $g>g_c$ the imaginary part of the quasiparticle spectrum is such that $\Sigma_{k}<0, \ \forall k$. }
  \label{fig:exact_diag_bis}
\end{figure*}

We notice that non-Hermitian extensions of classical Ising models have a long tradition in statistical physics~\cite{fisher1978yanglee,cardy1985conformal}, while the non-Hermitian TFIM has been much less explored (see however Ref.~\cite{hickey2013timeintegrated} for a realization in the context of  full-counting statistics transitions). Since this model is not PT symmetric~\cite{bender2015pt}, its many-body spectrum is in general complex. 

The non-Hermitian evolution at long times drive the state to the eigenstate of $\h_{\rm eff}$ with smaller imaginary part, the so called subradiant state, since
$\ket{\psi(t)} = \sum_{\alpha=1}^N \braket{\tilde{\alpha}|\psi(0)} e^{-i t E_\alpha} \ket{\alpha}$
where $\bra{\tilde{\alpha}}$/$\ket{\alpha}$ are the left/right eigenvectors of $\h_{\rm eff}$, $E_\alpha=\omega_\alpha -i \Gamma_\alpha$ are the corresponding complex eigenvalues and $N$ is the dimension of the Hilbert space.  
Remarkably, $\h_{\rm eff}$ undergoes a subradiance transition~\cite{dicke1954coherence,gross1982superradiance,celardo2009superradiance,auerbach2011superradiant,biella2013subradiant,
guerin2016subradiance,rotter2015areview,ashida2020nonhermitian} at a critical value of $g$ which is signaled by a non-analytic behavior of the eigenvalue with smaller imaginary part, $\Gamma_{\rm sub} =\min\left(\Gamma_\alpha\right)$ (see App.~\ref{app:spectral}). 
While for a non-Hermitian Hamiltonian such non-analyticity can occur sharply also in a finite size system in the following we will focus on those properties which are genuine many-body effects depending on the system size. The structure of the subradiant state when $g\rightarrow\infty$ can be immediately read from Eq.~(\ref{hdissdef}) and gives $ \ket{{\rm sub}} = \ket{T}$. In Figure~\ref{fig:Poverlap} (bottom panel) we plot the overlap $O_{{\rm sub}, T}$ between the subradiant state and the initial uncorrelated state $\ket{T}$ as a function of $g$, showing a monotonous decrease from $O_{{\rm sub}, T}\simeq 1$ (as $g$ is decreased) and a rapid drop around $g\simeq g_c$. Remarkably the behavior for $g>g_c$ matches with the one obtained from the full stochastic evolution by tracking the secondary peak in the probability distribution (see Fig.\ref{fig:Poverlap}).
A more clear indication of a critical behavior in the overlap can be seen by plotting the ratio between the probability of the subradiant state being inside or outside the Zeno subspace, as we do in Figure~\ref{fig:Sub_size2} (top panel)
\be
f=\frac{O_{{\rm sub}, T}}{1-O_{{\rm sub}, T}}.
\ee 
We see that $f<1$ for $g<g_c$, while for $g>g_c$ we have a large value of $f$, namely the subradiant state is essentially in the Zeno subspace. 
We stress that this sharp localization in the Hilbert space is highly nontrivial since the Zeno subspace dimension does not increase with the size of the system while the dimension of the full Hilbert space increases exponentially with $L$. 
Numerically we find $f\sim g^2$ for large $g$, from which we get $O_{{\rm sub}, T}=1-A/g^2$ at large value of $g$. Interestingly the overlap scales differently with system sizes depending on the value of $g$, as we show in Figure~\ref{fig:Sub_size2} (bottom panel), where an almost constant overlap is found for large $g$ while a power-law scaling is found for $g<g_c$ ($O_{{\rm sub},T}\sim L^{-\beta}$, with $\beta=0.049\pm 0.003$).

\section{Analytical approach}
The non-hermitian TFIM in Eq.~(\ref{hdissdef}) can be diagonalised exactly using the Jordan-Wigner representation of the spin operators into fermionic creation/annihilation operators and a generalised Bogoliubov transformation, as we discuss more in detail in App.~\ref{app:diag}. The final result, in terms of new fermionic operators 
$\hat{\bar{\gamma}}_k,\hat\gamma_k$ reads (fixing $J^x=1$)
\begin{equation}
\h_{\rm eff} = \sum_{k>0} \Lambda_k\left(\hat{\bar{\gamma}}_k\hat{\gamma}_k+\hat{\bar{\gamma}}_{-k}
\hat{\gamma}_{-k}-1\right) - i\frac{g}{g_c}L,
\end{equation}
where the quasiparticle spectrum is given by
\begin{equation}\label{eq:lambdak}
\Lambda_k=\pm 2\sqrt{1-(g/g_c)^2+2 ig/g_c\,\cos k}\equiv \Delta_k+i\Sigma_k
\end{equation}
and, following Ref.~\cite{lee2014heralded} we choose $\Sigma_k<0$ for all $k$, and we have introduced the critical coupling $g_c=4$. The behavior of this spectrum, plotted in Figure~\ref{fig:exact_diag_bis}, is particularly interesting around $k=\pi/2$: for $g<g_c$ the real-part is gapped, with a gap closing (in the thermodynamic limit $L\to\infty$) as $\Delta=\Delta_{k=\pi/2}\sim\sqrt{1-(g/g_c)^2}$ (with $\Sigma_{k=\pi/2}=0$); 
for $g>g_c$ the real-part of the energy spectrum remains pinned to zero ($\Delta_{k=\pi/2}=0$) while the imaginary part become non-vanishing as $\Sigma=\Sigma_{k=\pi/2}\sim - \sqrt{(g/g_c)^2-1}$. Right at the critical point, $g=g_c$, the spectrum around $k=\pm\pi/2$ acquires a very unusual power-law scaling
\be
\Lambda_{k}\sim \vert k\mp \pi/2\vert^{1/2}\left[\pm\mbox{sign}(k\mp\pi/2)-i\right].
\ee
This result has direct consequences in the structure of the two subradiant states merging at $g_c$, whose structure factor is strongly peaked at $k\simeq \pi/2$ before the transition (see App.~\ref{app:kmode}). Finally, from the quasiparticle spectrum~(\ref{eq:lambdak}) we can obtain the many-body eigenstates (see "Methods"), in particular the emerging subradiant state which has an (intensive) complex energy 
\be\label{eqn:subradiant}
\varepsilon_{\rm sub} = \frac{1}{2\pi}\int_{-\pi}^{0}\Lambda_k \ dk-i\frac{g}{ g_c}.
\ee
At weak coupling, $g/g_c\ll 1$, the subradiant eigenvalue has a real-part which does not depend on $g$ and a linearly growing imaginary part,  $\varepsilon_{\rm sub}\sim 1-i g/g_c$. At strong coupling instead, $g/g_c\gg 1$ we find from Eq.~\eqref{eqn:subradiant} a purely imaginary eigenvalue, which goes to zero as  $\varepsilon_{\rm sub}\sim - 2ig_c/g$, in agreement with our exact numerical calculations (see App.~\ref{app:diag}).

We conclude pointing out that the physics highlighted in this work would be completely invisible to the dynamics of the density matrix representing the statistical mixture of $N$ quantum states (here labelled by $n$) $
\rho(t) = \frac{1}{N}\sum_{n=1}^{N}\ket{\Psi_{n}(t)}\bra{\Psi_{n}(t)}$.
Given the Kraus operators definedin Eq.~\eqref{Measurements}, the density matrix $\rho(t)$ obeys the following Gorini-Kossakowski-Lindblad-Sudarshan (GKLS) master equation~\cite{petruccioneBook}
\be
\label{meq}
\partial_{t}\rho(t) = -\frac{i}{\hbar} \left[\h_{\rm eff},\rho(t) \right]+\sum_{i=1}^{L}\hat{L}_{i}\rho(t)\hat{L}_{i}^{\dagger},
\ee
where $\hat{L}_{i}=\sqrt{\gamma} \ \hat{\Pi}^z_{i+}$. The master equation~\eqref{meq} models a Quantum Ising model in presence of local dephasing processes taking place at a rate $\gamma$. It leads to the a fully depolarized steady-state density matrix ($\lim_{t\to\infty}\rho(t)=\rho_{\rm SS}$)
\be
\rho_{\rm SS} = \frac{1}{2^{L}} \bigotimes_{i=1}^{L} \mathbb{I}_{i}.
\ee
Therefore, the steady-state density matrix retains no information about the statistical features of the observables studied in this work. 
However, the analysis of the spectrum of the GKLS equation could reveal useful information associated to the typical time scale associated to the onset of the Zeno regime~\cite{snizhko2020quantumzeno}.

\section{Conclusion}

In this work we have shown that the many-body QZE in a quantum Ising chain, where strong continuous monitoring of the transverse magnetization leads to a complete freezing of the system, appears sharply upon increasing 
the ratio between measurement strength and coherent (entangling) interaction. Crucially, this transition is not encoded in the average dynamics, but it is rather controlled by rare fluctuations which we show are perfectly captured by looking at the no-click evolution. In the limit of continuous measurements this reduces to a non-hermitian TFIM which we show, using exact numerics and a mapping to fermions, to undergo a sharp subradiance transition characterized by a non-analytic behavior of the slowest decay mode and by a gap closing in the quasiparticle spectrum. This work establish a rather general important connection between many body QZE, subradiance transition in non-hermitian systems and rare fluctuation of the stochastic dynamics of a measured quantum many-body system. Future directions include the investigation of the role of spatially inhomogeneous measurements ($\psite<1$) leading to dilute disorder~\cite{harris1974upper,stinchcombe1981diluted,thomson2019griffiths,senthil1996higher}, different initial conditions and higher dimensionality.
The study of a model where the unmeasured dynamics is governed by an interacting model represents a further intriguing research direction.

\acknowledgements{We thanks useful discussions with L. Mazza and R. Fazio.
The  computations  were  performed  on  the  Coll\'ege de France IPH computer cluster.  We acknowledge support from  the ANR grant “NonEQuMat”(ANR-19-CE47-0001). AB acknowledges funding by LabEx PALM (ANR-10-LABX-0039-PALM).}


\section*{APPENDICES}
\appendix

\section{Diagonalization of the non-Hermitian quantum Ising model}
\label{app:diag}
We consider the non-Hermitian Transverse Field Ising Model (TFIM) introduced in the main text, whose Hamiltonian reads
\begin{equation}\label{eqn:nh_ising}
\h_{\rm eff} = \h -i \hat{\Gamma}= J^x\sum_{i=1}^{L-1}\ssx_i\ssx_{i+1}-i\frac\gamma2\sum_{i=1}^L\frac{\ssz_i+1}{2}
\end{equation}
and we take periodic boundary conditions for the spin $\sigma^x_{L+1}=\sigma^x_1$. Using the Jordan-Wigner transformation, which maps spin-$1/2$ operators to fermions through the identities
\begin{eqnarray}
\label{eqn:JWsigmax}\sigma^x_j  &=& \hat K_j\left(\hat c_j+\hat c^{\dagger}_j\right),\\
\label{eqn:JWsigmaz}\sigma^z_j &=& 1-2 \hat c^{\dagger}_j\hat c_j,
\end{eqnarray}
where the so called {\itshape string} operator $\hat K_j$, defined as
\be
\hat K_j=\left[\prod_{l<j} \left(1-2\hat c^{\dagger}_l\hat c_l\right)\right]\,,
\ee
ensures the proper fermionic algebra. 
We can rewrite Eq.~(\ref{eqn:nh_ising}) as
\begin{eqnarray}\label{eqn:fermionized}
\h_{\rm eff}  &&= J^x\sum_{i=1}^{L-1}\left(\hat c^{\dagger}_i\hat c_{i+1}+\hat c^{\dagger}_i\,\hat c^{\dagger}_{i+1}+{\rm H.c.}\right) \cr
&&-i\frac{\gamma}{2}\sum_{i=1}^L\,\left(1-\hat c^{\dagger}_i\hat c_i\right)\nonumber\\
&&+(-1)^{N_F}J^x\left(\hat c^{\dagger}_L\hat c_{1}+\hat c^{\dagger}_L\,\hat c^{\dagger}_{1}+{\rm H.c.}\right).
\end{eqnarray}
In the last expression $N_F=\sum_{i=1}^L \hat c^{\dagger}_i \hat c_{i}$ is the total number of fermions and $(-1)^{N_F}$ the associated parity. We notice that in the Hamiltonian case (purely real transverse field) the fermionic parity is conserved and as such the last term is just a phase that can be handled by proper choice of boundary conditions, i.e. by posing $\hat c_{L+1}=(-1)^{N_F+1}\hat c_1$.
In the non-Hermitian case this is in general not the case, however one can show that starting from an initial state which is eigenstate of the fermionic parity (as it is our case here) the latter will remain conserved throughout the evolution. Therefore we fixed the parity to be even (as we start from the $N_F=0$ sector), which corresponds to anti-periodic boundary conditions (ABC). Going to Fourier space, and fixing the values of $k$ appropriate for ABC, we obtain
\ba
 \h_{\rm eff} &=&\sum_k\,\varepsilon_k\,\hat{c}^{\dagger}_k\hat{c}_k -i\frac{\gamma}{2}L\cr
 &&   +J\sum_k\,i\sin k\left(\hat{c}^{\dagger}_k\,
 \hat{c}^{\dagger}_{-k}- \hat{c}_{-k}\hat{c}_k\right)
\ea
with $\varepsilon_k=i\gamma/2+2J\cos k$. We notice that last term on the right hand side, deriving from Eq.~(\ref{eqn:fermionized}) which is crucial to ensure the non-Hermitian Hamiltonian has a positive definite imaginary part.
Restricting the sum to positive values of $k$ and introducing Nambu-Gorkov spinor notation $\Psi_k=\left(\hat{c}_k, \hat{c}^\dagger_{-k}\right)^{\rm T}$ we get 
$
\h_{\rm eff} = \sum_{k>0} \bar{\Psi}_k h_k \Psi_k -i\frac{\gamma}{4}L
$
where we defined the generalized Bogoliubov matrix
\be
\label{bogmat}
h_k = 
\begin{pmatrix}
\varepsilon_k & -i\eta_k \\
i\eta_k  & -\varepsilon_k
\end{pmatrix},
\ee
where $\eta_k=-2J^x\sin k$. We can then diagonalise the matrix $h_k$ and introduce new fermionic operators 
$\hat{\bar{\gamma}}_k,\hat{\gamma}_k$ to obtain
\begin{equation}
\h_{\rm eff} = \sum_{k>0} \Lambda_k\left(\hat{\bar{\gamma}}_k\hat{\gamma}_k+\hat{\bar{\gamma}}_{-k}
\hat{\gamma}_{-k}-1\right) -i\frac{\gamma}{4}L,
\end{equation}
where the quasiparticle spectrum reads
\ba
\label{lambdaApp}
\Lambda_k&=&\pm 2J^x\sqrt{1-(g/g_c)^2+2ig/g_c\,\cos k} \cr
&&\cr
&\equiv& \Delta_k + i \Sigma_k
\ea
and we have introduced the critical coupling $g_c=4$. 
Once the Hamiltonian is diagonalised we can write down the vacuum state of $\h_{\rm eff}$
\begin{equation}
\vert \rm{vac}\rangle = \prod_{k>0}\hat{\gamma}_{-k}\hat{\gamma}_k\vert 0\rangle
\end{equation}
with intensive energy $\varepsilon_{\rm vac}=E_{\rm vac}/L=-\frac{1}{L}\sum_{k>0} \Lambda_k-i\frac{\gamma}{4}$ which reads (in the thermodynamic limit)
\begin{eqnarray}
\varepsilon_{\rm vac}/2J^x&=&-\int_0^{\pi}\frac{dk}{2\pi}\sqrt{1-(g/g_c)^2+2ig/g_c\,\cos k}\cr
&&\cr
&&- i\frac{g}{2g_c}.
\end{eqnarray}
Similarly, any excited state of the form 
\begin{equation}
\vert \left\{ n_k\right\}\rangle = \prod_{k}\left(\hat{\bar{\gamma}}_k\right)^{n_k}\vert \rm vac\rangle
\end{equation}
with $n_k=0,1$ such that $\sum_k n_k=N_F$, given the even parity sector, has eigenvalue
\begin{equation}
E( \left\{ n_k\right\})=E_{\rm vac}+\sum_{k}n_k \Lambda_k.
\end{equation}
For example, the state $\ket{T}=\bigotimes_{i=1}^L\ket{-}_i$ (that is the exact subradiant state of $\h_{\rm eff}$ in the $g/g_c\to\infty$ limit) can be written in this fermionic language as the maximally occupied state, with $n_k=1$ for all $k$. It has eigenvalue 
\begin{eqnarray}
\varepsilon_{\rm sub}/2J^x&=&\int_{-\pi}^{0}\frac{dk}{2\pi} \sqrt{1-(g/g_c)^2+2 i g/g_c\,\cos k}\cr
&&\cr
&&-i\frac{g}{2 g_c}.
\end{eqnarray}

\section{Excess entanglement}
\label{app:miee}

\begin{figure}[h!]
\centering
\includegraphics[width=0.9\columnwidth]{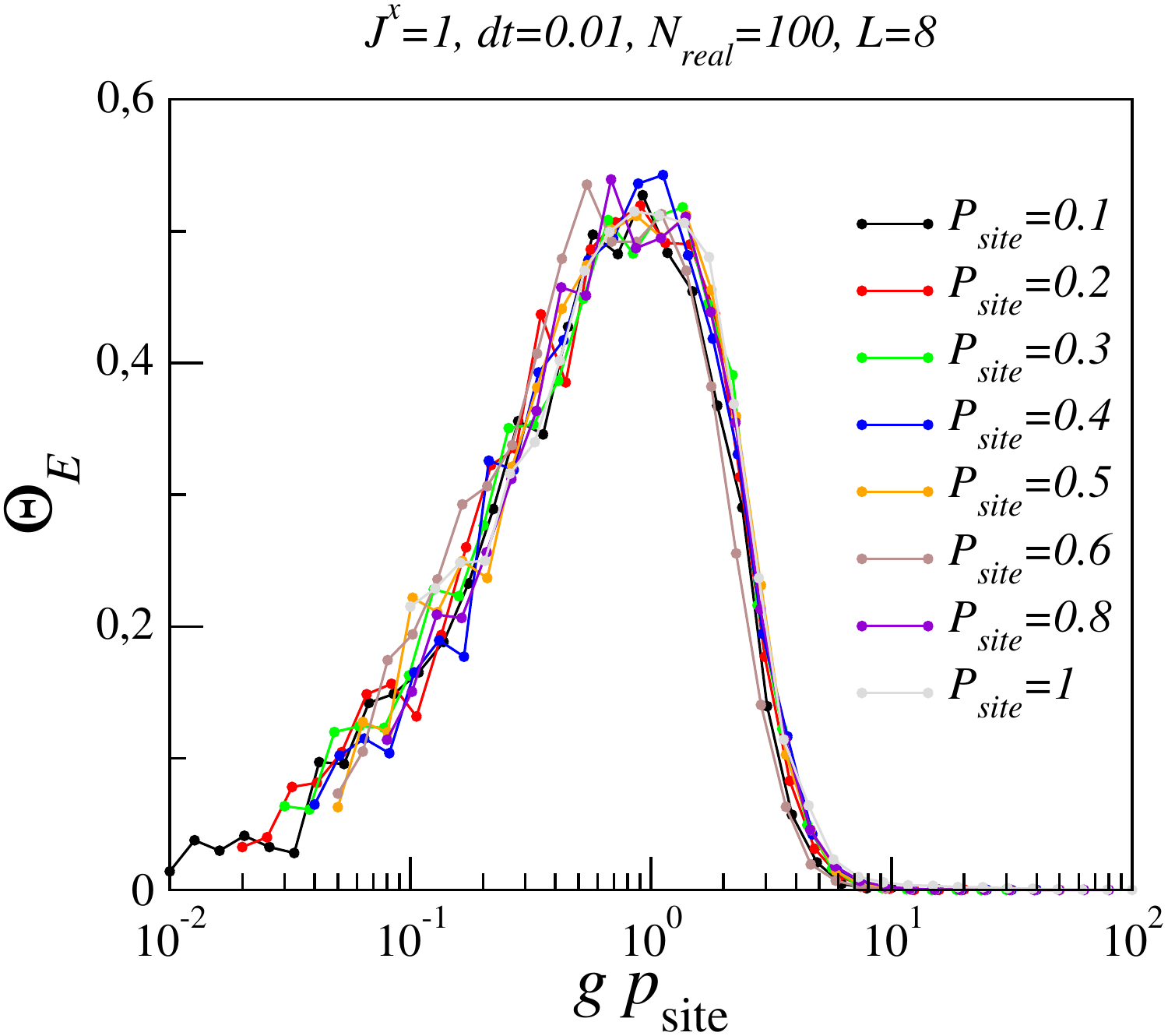} 
\caption{Measurement-induced entanglement $\mie$ as a function of $g \ \psite$ for several values of $\psite$.
Here we set $N_{\rm real}=100, dt=10^{-2},L=8,J^x=1$.}
  \label{fig:EEscalingApp}
\end{figure}

As discussed in the main text, we quantify the {\it excess entanglement} via the quantity $\mie$ defined in Eq.~\eqref{mie}. In this appendix we show how $\Theta_{E}$ depends on $\psite$.
In Fig.\ref{fig:EEscalingApp} we show  $\Theta_{E}$ for several different values of $\psite$. 
Curves collapse when plotted against $g \ \psite $ [with $g$ defined in Eq.~\eqref{cpar}].
This result shows that that for this specific quantity the control parameter is given by 
\be
g \ \psite =\frac{p_{\rm tot}}{J^x \ dt},
\ee
where $p_{\rm tot}=\psite \ \pmeas$, indicating that the spatial density and the strength of the measurements play an equivalent role in the generation of the excess entanglement.

\section{Validity of the non-Hermitian effective Hamiltonian}
\label{app:heff_validity}
In this appendix we test the validity of the non-Hermitian effective Hamiltonian $\h_{\rm eff}$.
To this purpose we compare the {\it full} no-click evolution  with the one generated by the $\h_{\rm eff}$ [Eq.~\eqref{hdissdef}] for different values of $g$. 

\begin{figure}[h!]
\centering
\includegraphics[width=0.9\columnwidth]{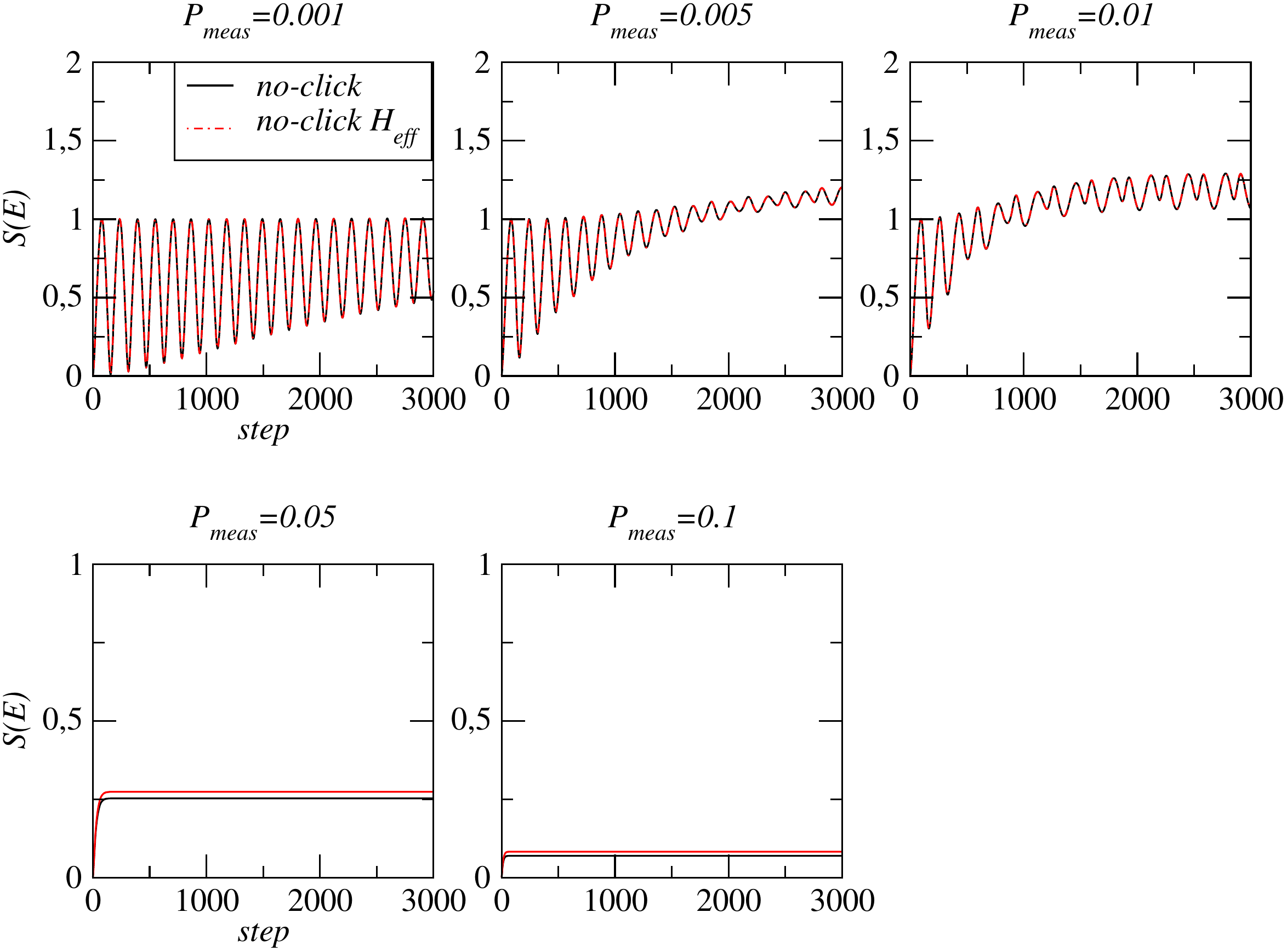}
\caption{Comparison between the {\it full} no-click evolution with the one generated by the $\h_{\rm eff}$ [Eq.~\eqref{hdissdef}] for different values of $\pmeas=\gamma dt$. Here we set $dt=10^{-2},L=8,J^x=1$.}
  \label{fig:Heff_validity}
\end{figure}

As we can see from Fig.~\ref{fig:Heff_validity} we dynamics induced by $\h_{\rm eff}$ reproduce exactly the no-click evolution of the entanglement entropy $S_E$ for $g\lesssim1$. After this threshold the results for the entanglement entropy slightly differ quantitatively but behaves qualitatively in the same way. 

%

\section{Onset of the Zeno effect}
\label{app:zeno}

In this section we show appearance of the Zeno dynamics as $g$ is increased.
The onset of the SZ phase can be inspected monitoring the probability distribution of the local magnetization $P(m^z)$ which represents the probability of having a certain value of the local magnetization during the dynamics $m^z=\bra{\psi(t)}\ssz_i\ket{\psi(t)}, \forall i$.
For $g\lesssim 1 $ the distribution $P(m^z)$ explores almost uniformly all the possible values of the magnetization with two peaks progressively emerging at $m^z_-$ and $m^z_+$, located respectively at $m_z<0$ and $m_z\simeq1$.
For $g\gtrsim 1$ the structure of the probability distribution change dramatically. 
Beside the {\it usual} peak at $m^z\approx 1$, the second peak at $m^z_-$ sharper and moves toward $m^z=-1$ as $g$ is increased. Finally, for $g\gg1$ the distribution is a delta function peaked at $m^z=-1$. In the top left panel we also show the behaviour of $m_z^-$ as a function of $g$.

\begin{figure}[h!]
\centering
\includegraphics[width=0.9\columnwidth]{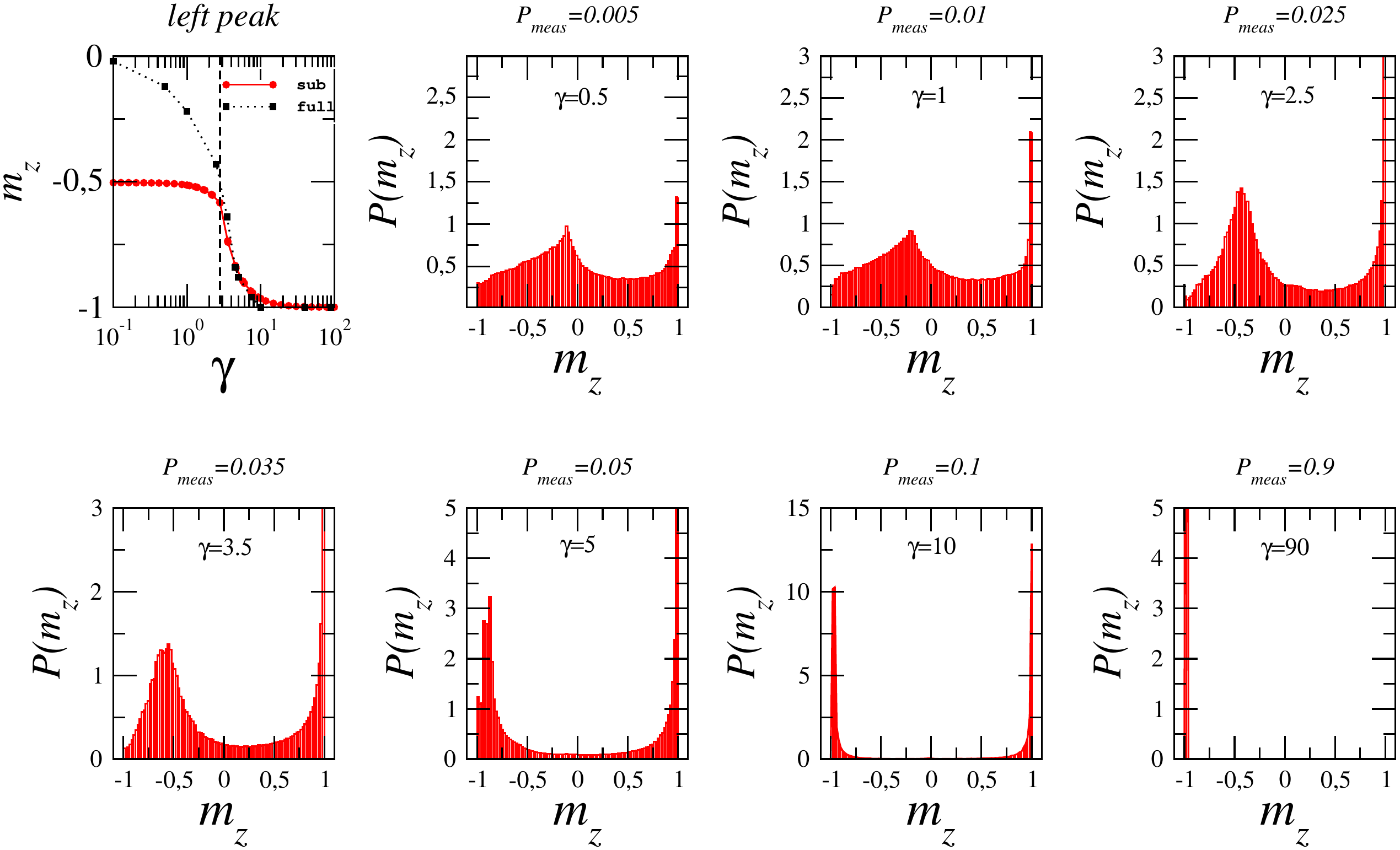}
\caption{Probability distribution of the local magnetization $P(m^z)$ for different values of $g$. In the top left panel we show the position of $m_z^-$, the leftmost peak of $P(m^z)$.  We also compare it with magnetization of the subradiant state of $\h_{\rm eff}$. Here we set $\psite=1, dt=10^{-2},L=8,J^x=1, N_{\rm real}=100$.}
  \label{fig:ZenoOnset}
\end{figure}

We stress that the value of $m^z$ unambiguously determine the quantum state since the local single-site density matrix can be parametrized as
$
\hat{\rho}_{\rm loc} = \frac12 \left(\mathbb{I} + m^z \ssz \right).
$
It is clear that the only free parameter is $m^z$, therefore the structure of $P(m^z)$ directly gives information about the onset of the Zeno effect (i.e. the system state is mostly frozen next to one of the measurement eigenstates, yet rarely performs quantum jumps between them). 
In Fig.~\ref{fig:ZenoOnset} we show the probability distribution of the local magnetization $P(m^z)$ which represents the probability of having a certain value of the local magnetization during the dynamics $m^z=\bra{\psi(t)}\ssz_i\ket{\psi(t)}, \forall i$.
In the top left panel we compare the position of the peak at $m^z_-$ with the magnetization of the subradiant state $\bra{\rm sub}\hat{\Sigma}^z\ket{\rm sub}/L$. The two curves show a perfect agreement for $g>4$.

\section{Spectral signatures of the superradiance transition}
\label{app:spectral}
\begin{figure}[h!]
\centering
\includegraphics[width=0.9\columnwidth]{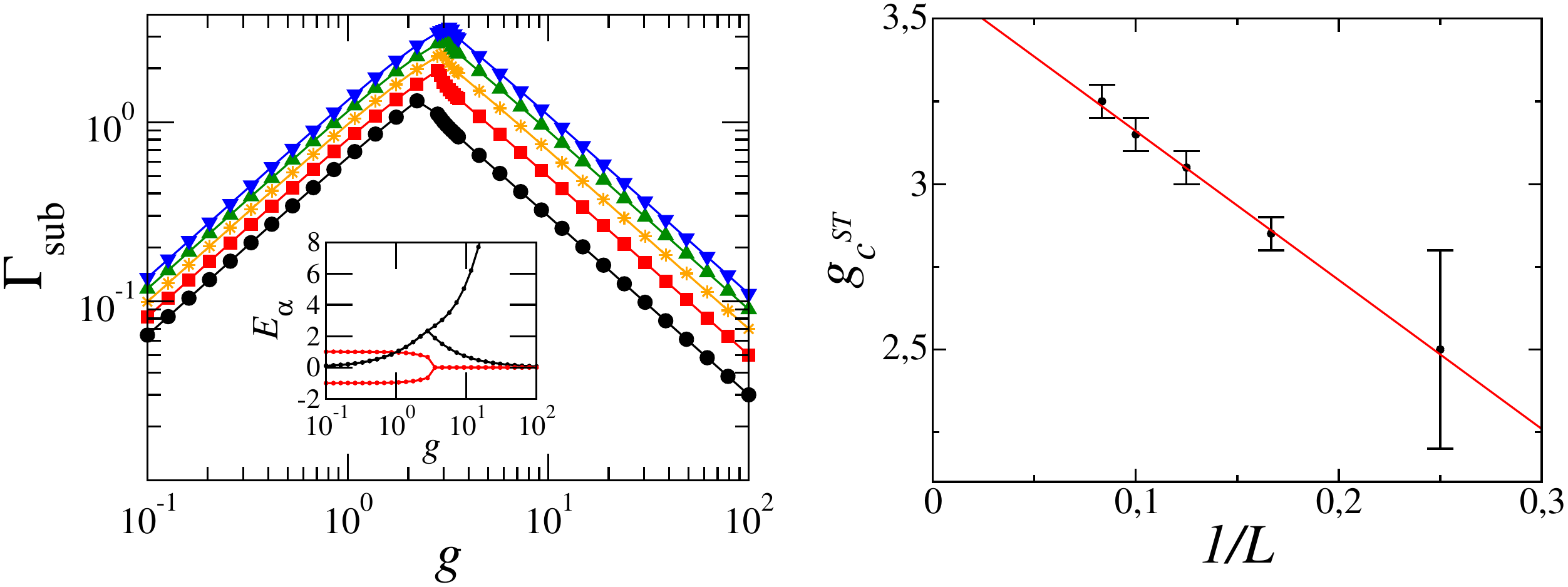}
\caption{Subradiant Transition in the Non-Hermitian Hamiltonian. Left panel: Subradiant width $\Gamma_{\rm sub}$ as a function of $g$. (Inset: Real and Imaginary part of the subradiand and its anti-subradiant partner as a function of $g$). Right panel: finite size scaling of the critical point $g_c^{ST}$ computed from the behavior of $\Gamma_{\rm sub}$ in the left panel. Here we set $dt=10^{-2},J^x=1$.}
  \label{fig:Sub_size}
\end{figure}

While in the main text we showed the emergence of the subradiance transition computing the overlap with the Zeno subspace, here we compute the spectral properties of the non-Hermitian Hamiltonian $\h_{\rm eff}$ [Eq.~\eqref{hdissdef}]. 
In particular, In Fig.~\ref{fig:Sub_size} (left panel) we show the behavior of the width of the subradiand state as a function of the dimensionless coupling $g$.
The real part of the subradiant eigenvalue vanishes for $g=g_c^{ST}$, where it merges with an anti-subradiant partner, and remains pinned there for $g>g^{ST}_{c}$ (left panel, inset). The transition to the subradiant regime is sharp also at finite size and the critical value of $g$ slowly shifts with size. A finite size scaling analysis of the critical point $g_c^{ST}$ (bottom panel, inset) gives $g_c^{ST}=\mathcal{A} + \mathcal{B} \ L^{-1}$ with $\mathcal{A}=3.61\pm0.02$ and $\mathcal{B}=4.5\pm0.1$. We can thus infer the behavior in the thermodynamic limit $(\lim_{L\to\infty}g_c^{ST} = \mathcal{A}$).

\section{$k$-mode occupancy}
\label{app:kmode}
Here we show the $k$-mode occupancy of the subradiant and anti-subradiant state (the many-body state that becomes degenerate with the subradiant at the transition) defined as
\begin{figure}[h!]
\centering
\includegraphics[width=0.7\columnwidth]{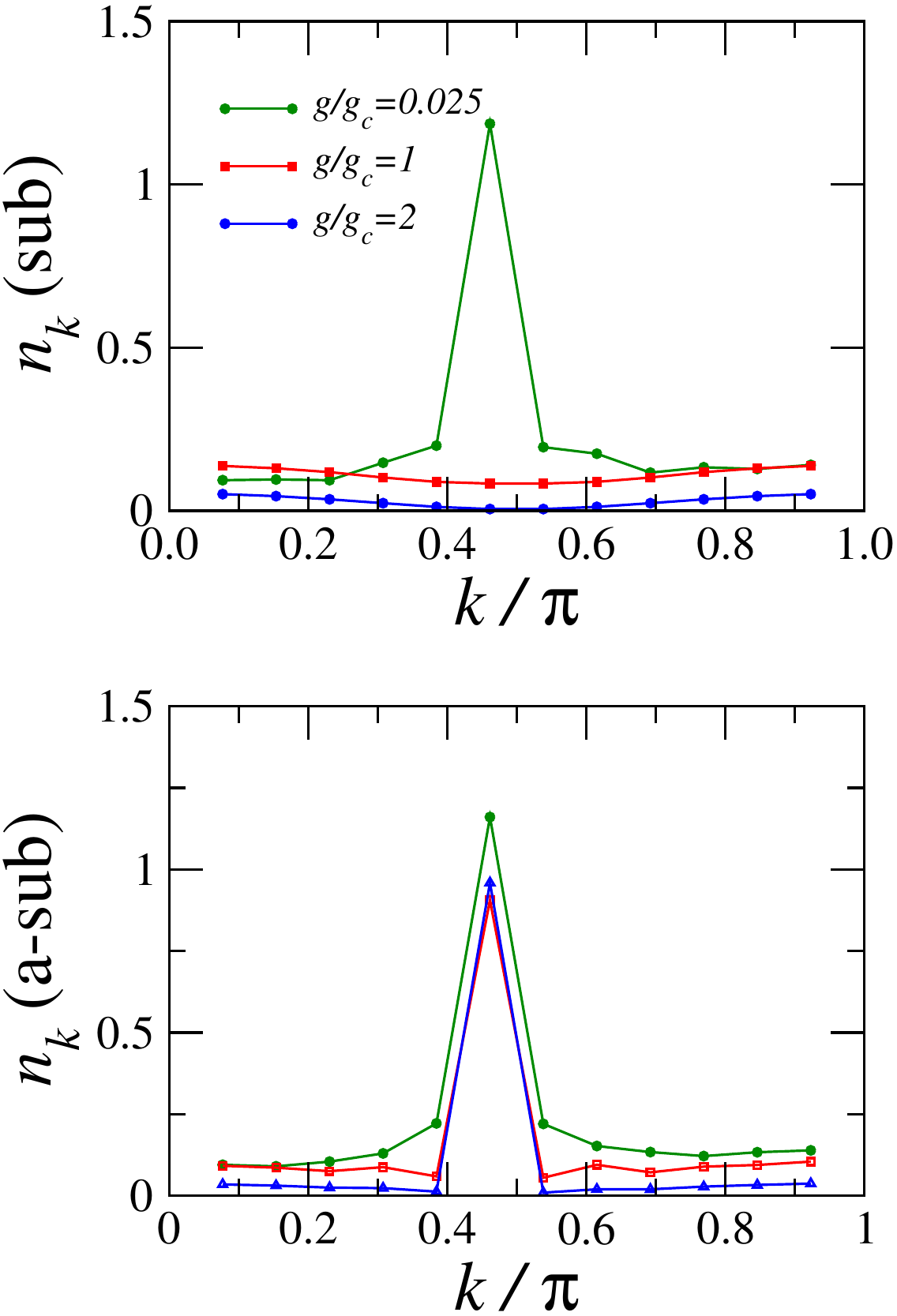}
\caption{Population in $k$-space [see Eq.\eqref{kmodeocc}] of the subradiant (top panel) and anti-subradiant (bottom panel) state. Here we set $J^x=1$ and vary $\gamma$ in order to change the dimensionless coupling $g$.}
  \label{fig:kmodes}
\end{figure}
\ba
\label{kmodeocc}
\braket{\hat{n}_{k}} &=& \braket{\ssp_k\ssm_k} \\
&&\cr
&=&\frac{2}{L+1}\sum_{i,j=1}^L \sin(k i) \sin(k j) \braket{\ssp_i \ssm_j}\nonumber,
\ea
with $k=n \pi/(L+1)$, $n=1,\dots,L$ as imposed by open boundary conditions and where we used the Fourier transform of the spin operators
\be
\ssp_k=\sqrt{\frac{2}{L+1}}\sum_{i=1}^L \sin(k i)  \ \ssp_i.
\ee 
The result is shown in Fig.~\ref{fig:kmodes}. The anti-subradiant state (bottom panel) remains mainly populates the $k=\pi/2$ mode of the chain as $g$ is varied across the transition. The superradiant state (top panel) is peaked around $k=\pi/2$ for $g/g_c<1$. After the transition the $k$-mode distribution is sharply depleted since $\ket{\rm sub}\to\ket{T}$ as $g/g_c\to\infty$.

\printbibliography

\blank

\end{document}